\documentclass[a4paper,12pt]{article}
\usepackage{epsfig}
\def\fnote#1{\footnote}

\def\square{\hbox{\vrule\vbox{\hrule\phantom{o}\hrule}\vrule}}

\newtheorem{teo}{Theorem}

\newcommand{\be}{\begin{equation}}

\newcommand{\ee}{\end{equation}}
\newcommand{\beqa}{\begin{eqnarray}}
\newcommand{\eeqa}{\end{eqnarray}}
\newcommand{\ba}{\begin{array}}
\newcommand{\bal}{\begin{array}{l}}
\newcommand{\ea}{\end{array}}
\newcommand{\bt}{\begin{teo}}
\newcommand{\et}{\end{teo}}

\begin{document}

\begin{center}
\epsfig{file=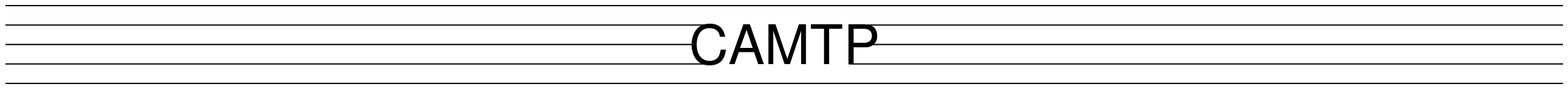,height=8mm,width=\textwidth}\\[2mm]
\end{center}

\begin{flushright}
Preprint CAMTP/98-7\\
July 1998\\
Revised December 1998\\
\end{flushright}

\vskip 0.5 truecm
\begin{center}
\Large
{\bf   On Urabe's criteria of isochronicity}\\
\vspace{0.25in}
\normalsize
Marko Robnik\footnote{e--mail: robnik@uni-mb.si}
 and Valery G. Romanovski\dag\footnote{e--mail: math@micro.rei.minsk.by}

\vspace{0.3in}
Center for Applied Mathematics and Theoretical Physics,\\
University of Maribor, Krekova 2, SI-2000 Maribor, Slovenia\\
\dag Belarusian State University of Informatics and Radioelectronics\\
P. Brovka 6, Minsk 220027, Belarus\\
\end{center}

\vspace{0.3in}

\normalsize
\noindent
{\bf Abstract.}
We give a short proof of  Urabe's criteria for the isochronicity of periodical
solutions of the equation ${\ddot x }+g(x)=0.$ We show that apart from
the harmonic oscillator there exists a large family of isochronous
potentials which must be all non-polynomial and not symmetric (even
function of the coordinate $x$).

\vspace{0.6in}

PACS numbers: 46.10.+z, 95.10.Ce\\
AMS classification scheme numbers:  34C05\\

Published in {\bf Journal of Physics A: Mathematical and General}\\
Vol. 32 (1999) 1279-1283
\normalsize
\vspace{0.1in}

\newpage

We consider
a system of differential equations of the form
\be   \label{sl}
\begin{array}{l}
{\dot x }  = y,\\
{\dot y } = -g(x),
\end{array}
\ee
where we suppose
\be \label{cond}
g(x)\in C(a,b), \ x g(x)>0 \mbox{ for}\ x\ne 0, g(0)=0 \mbox{ and} \
g'(0)=k \not= 0.
\ee
Denoting
$$
  U(x)=\int_0^x g(s)ds
$$
we obtain the first integral in the form "kinetic energy+potential energy",
i.e. in the form
\be \label{e}
  H(x,y) \stackrel{def}{=}  {y^2 \over 2}+U(x) = E,
\ee

such that $H(x,y)$ is the Hamiltonian and (\ref{sl}) are the Hamilton
equations of the motion of our system \cite{ll}.

It is well known that
any solution near the origin
oscillates around  $x=0, y=0$ with a bounded period, i.e.
system (\ref{sl}) has a center in the origin.
The problem arises then to
determine whether the period of oscillations is constant
for all solutions near the origin.
A center with such property is called $isochronous$.
At present the problem of isochronicity is of renewed interest
(see, for example, \cite{ggmm} for current references).

It was shown in \cite{cj} that
if $g(x)$ is a polynomial, then system (\ref{sl}) cannot have
an isochronous center, except when $g(x)$ is linear $g(x)=kx$,
in which case $k=(2\pi/\tau)^2$, where $\tau$ is the period
of oscillations. If $g(x)$ is not exactly linear, then still
the period of oscillations infinitesimally close to the
origin is also equal to $\tau$.

In the present Letter we give a simple short proof
of the following Urabe's  criteria  \cite{ur}
of isochronicity of the center of system
(\ref{sl}).
\bt  \label{u1}
  When $g(x)$ is continuous, the necessary and sufficient condition that
  $g(x)\in C^1(a,b)$  and system  (\ref{sl}) has an isochronous center in the
  origin, is that, in the neighbourhood of $x=y=0$ by the transformation
  \be \label{X2}
    \frac 12 X^2=  U(x),
  \ee
  where $X/x>0$ for $x\ne 0,$ $g(x)$ is expressed as
  \be \label{eg}
    g(x)=g[x(X)]\stackrel{def}{=}h(X)=\frac {2\pi}\tau {X \over {1+S(X)}},
  \ee
  where $S(X)$ is an arbitrary continuous  odd function and $\tau$
  is the period of the oscillations.
\et

First in \cite{ur}
Urabe proved the criteria in the case when $g(x)$ is an analytic function.
For function $g(x)\in C^1$ he got a  more complicated criteria with
the function $h(X)$ of the form
$$
  h(X)=\frac {2\pi}\tau {X \over {1+S(X)+R(X)}},
  $$
  where $S(X)$ is an  odd and $R(X)$ is an even  continuous
function (see \cite{ur}).
Then in \cite{ur2} he showed that if $g(x)\in C^1(a,b)$
then necessarily $R(X)\equiv 0.$

Note that in the statement of the theorem Urabe demands the additional
property
$$
  S(0)=0,\;\;\; XS(X)\in C^1,
$$
but every continuous odd function has the property $S(0)=0$, and
the second one is not essential for our proof. We have also required
$g(x)$ to be smooth in a neighbourhood of $x=0$ (as in the original
work by Urabe \cite{ur}, but in fact it is sufficient  for our
reasoning if $g(x)$ is continuous in a neighbourhood of the origin
and differentiable at $x=0$.

Our proof of the Theorem 1 is based on the following
criteria, which  for the first time appears, apparently, in
Landau and Pyatigorsky \cite{lp} and which later
was rederived by Keller \cite{k62,k76}  (who also considered
some connected problems, in particular, the case of non-monotonic potential).
For convenience of the reader we present the criteria with the  proof,
which stems from the books  \cite{ll,lp}, here.
\bt \label{l1}
  When $g(x)$ is continuous and the conditions (\ref{cond}) hold,
  system (\ref{sl}) has an isochronous center of the period $\tau$
  at the origin
  if and only if
  \be  \label{cl}
     x_2(U)-x_1(U)={{\sqrt 2 \tau} \over \pi} \sqrt U ,
  \ee
  for $U\in (0,U_0)$,
  where $x_1(U)$ is the inverse function to $U(x)$ for $x\in (a,0)$
  and $x_2(U)$ is the inverse function to $U(x)$ for  $x\in (0,b)$.
\et
{\it Proof}. First we note that due to (\ref{cond}) the functions
$x_1(U), x_2(U)$ are defined and $x_1(U), x_2(U)\in C^1(0,U_0)$
with a $U_0>0$. Denote by $T(E)$ the period of the orbit of (\ref{sl})
corresponding to the value of energy $E$.
Then we have \cite{ll}
\be \label{te}
  T(E)= \sqrt 2
  \int_0^E \left[ {{dx_2(U)}\over {dU}}-{{dx_1(U)}\over {dU} }
  \right] {{dU} \over \sqrt{E-U}}.
\ee
Dividing both sides of this equation by $\sqrt{\alpha-E} $, where
$\alpha $  is a parameter, integrating with respect to $E$ from 0
to $\alpha$ and putting U in place of $\alpha$
(see \cite{ll} for detail) one gets
$$
      x_2(U)-x_1(U)={1 \over {\sqrt 2 \pi}}
      \int_0^U {T(E)dE \over \sqrt{U-E}}.
$$
In the case when $T(E)\equiv \tau$ that yields (\ref{cl}).

To prove that (\ref{cl}) is the sufficient condition of isochronicity
we note  that (\ref{cl}) implies
  $$
     x_2'(U)-x_1'(U)={{\sqrt 2 \tau} \over {2\pi \sqrt U}}.
  $$
Substituting this expression
into (\ref{te}) and integrating we get $T(E)\equiv \tau$.
\square

\newtheorem{cor}{Corollary}
As an  immediate consequence we get the following proposition
proved earlier in \cite{ur}.
\begin{cor}
   If $g(x)\in C^1(a,b)$ is an odd function,
then the origin is an isochronous
   center iff $g(x)=(2\pi/\tau)^2 x.$
\end{cor}

In other words, if the potential (energy) $U(x)$ is an even
function of position $x$ then the only isochronous system
is the harmonic oscillator given above.

{\it Proof of theorem \ref{u1}}.
Let us suppose that the system (\ref{sl}) has an isochronous center.
Then due to theorem \ref{l1} the relation (\ref{cl}) holds and we get
$$
     x_2(U)-{{\sqrt 2 \tau} \over {2 \pi}} \sqrt U=
     x_1+{{\sqrt 2 \tau} \over {2 \pi}} \sqrt U \stackrel{def}{=} f(U).
$$
Therefore
\begin{eqnarray} \label{x2}
   x_2'(U)= {{\tau} \over {2 \sqrt 2 \pi \sqrt U}}+f'(U),\\
   \label{x1}
    x_1'(U)= {-{\tau} \over {2 \sqrt 2 \pi \sqrt U}}+f'(U).
\end{eqnarray}

Taking derivative in the both parts of (\ref{cl}) with respect to $x$
we get for $x<0$
$$
  x_2'(U)U'- 1=  {{\sqrt 2 \tau} \over {2  \pi \sqrt U}}U'.
$$
Therefore, using (\ref{x2}) we obtain
\be \label{u2}
 U'=  {{2\pi} \over \tau}
 {-{ \sqrt{2 U}} \over {1-{{2\pi}\over \tau}
 \sqrt {2U}f'(U)}}.
\ee
Similarly, for $x>0$ we get from (\ref{x1})
\be \label{u11}
 U'=  {{2\pi} \over \tau}
 {{ \sqrt{2 U}} \over {1+{{2\pi}\over \tau}
 \sqrt {2U}f'(U)}}.
\ee
 Therefore     function $g(x)$ can be expressed in the form (\ref{eg}).

Now it remains to show that
$$
  S(X)={{2\pi}\over \tau} X f'(X^2)
$$
is a continuous function.
Obviously, it is true if $X\ne 0$.

For $X=x=0$ we have the situation as follows.
First note that (\ref{cond})  and (\ref{cl}) yield
$$
  U=\frac {2\pi^2}{\tau^2}x^2+o(x^2).
$$
Then  for $x,X>0$ from (\ref{u11}) we get
$$
   S(X)= {{2\pi}\over \tau}
 \sqrt {2U}f'(U)= {{{2\pi^2}\over \tau}
 \sqrt {2U} \over {U'}}-1={x\sqrt{1+o(1)}\over {x+o(x)}}-1.
$$
Therefore
$$
  \lim_{X\to 0+}S(X)=0.
$$
For $x,X<0$ (\ref{u2}) yields
$$
   S(X)=- {{2\pi}\over \tau}
 \sqrt {2U}f'(U)=- {{{2\pi^2}\over \tau}
 \sqrt {2U} \over {U'}}-1=-{|x|\sqrt{1+o(1)}\over {x+o(x)}}-1.
$$
It means $
  \lim_{X\to 0-}S(X)=0
$
and, hence, $S(X)$ is continuous at zero.

 Let us prove that (\ref{eg}) is also the  sufficient  condition
 of isochronicity.
 For $x>0$ we can write (\ref{eg}) in the form
 $$
    {{dU} \over {dx}}=\frac {2\pi}\tau {X \over {1+S(X)}} =
          \frac {2\pi}\tau {\sqrt{2U} \over {1+S(\sqrt{2U})}}.
 $$
 Integrating this  equation we get
$$
  x_2(U)={\tau \over {2\pi}}(\sqrt{2U}+\int_0^{\sqrt{2U}} S(z) dz).
$$
Similarly, for $x<0$ we obtain
$$
  x_1(U)={\tau \over {2\pi}}(-\sqrt{2U}+\int_0^{\sqrt{2U}} S(z) dz).
$$
Due to the condition of the theorem $S(z)$ is a continuous function,
and, hence, the integral is convergent.
Therefore  (\ref{cl}) holds, i.e. the system has an isochronous
center in the origin.
\square

In conclusion, we have proven that the Hamiltonian (\ref{e})
has the isochronous center iff the condition (\ref{cl}) is satisfied.
In case of a symmetric potential $U(x)$ (even function of $x$)
the only solution is the harmonic oscillator. If $U(x)$ is
not symmetric (even), other solutions might be possible. However,
for any polynomial $U(x)$ (and $g(x) = U'(x)$), the harmonic
potential is still the only solution \cite{cj}. Thus, other
nontrivial isochronic potentials can be invented by taking
an analytic but not polynomial and not even function $U(x)$,
in agreement with Urabe's criteria (\ref{eg})  of Theorem 1,
which we have shown to be equivalent to (\ref{cl}). These
criteria allow still for a quite large family of isochronous
potentials  $U(x)$ and we can construct such potentials analytically.
Indeed, differentiating the both sides of the equality
(\ref{X2})
and taking into account (\ref{eg})  we get in the case of isochronous center
$$
  X{{dX}\over {dx}}=g(x)=
  \frac {2\pi}\tau {X \over {1+S(X)}}.
$$
Hence, we obtain the next formula, which for the first time appears
in  \cite{ur}
\begin{equation} \label{xx}
      x=  \frac \tau {2\pi}\int_0^X \left( 1+S(u) \right) du.
\end{equation}
This formula together with (\ref{eg})   is  a  tool  to
construct
isochronous potentials. Taking $S(X)=X$ Urabe got
$$
  g(x)=\frac {2\pi}\tau [1-(1+\frac {4\pi}\tau x)^{-\frac 12}],
$$
hence, the corresponding isochronous potential is
\begin{equation}   \label{exu}
U(x)=  1 +\frac {2\pi}\tau x-
   \sqrt{1+\frac {4\pi}\tau x}.
\end{equation}
where $-\frac \tau{4\pi}< x< {{3 \tau}\over {4\pi} }$,
i.e. the potential is an analytic function defined
on a finite segment of real axis. Here, in the calculation,
we have chosen the (negative) sign such that $g(x=0)=0$ is
obeyed.

Let now
$$
  S(X)=\frac 2\pi {\rm arctg}\, X.
$$
Then (\ref{xx}) yields
$$
  x=\frac \tau{2\pi} X + \frac \tau{\pi^2} X {\rm arctg}\, X -\frac
\tau{2\pi^2}
  \log (X^2+1).
$$
Obviously, $x(X)$ is strictly increasing on ${\bf R}$
and $x(0)=0, \ x({\bf R})={\bf R}.$ Therefore,
$$
  g(x)= \frac {2\pi}\tau  {{X(x)} \over {1+\frac 2\pi {\rm arctg}\, (X(x)) }}
$$
is defined for all $x\in {\bf R}$, positive for
$x>0$ and negative for $x<0.$ Hence, the corresponding
potential $U(x)$ is an analytic function defined on the whole
real axis with the only one minimum in the origin.
One can construct this potential at least in the form of power series.
However, the potential is not an entire function.
As we have mentioned above it was shown in \cite{cj}  (in fact, it is an
immediate consequence of formula (\ref{cl})),
that the only polynomial isochronous potential is the quadratic one.
We also see that there are analytic potentials defined on whole
real axis. Thus the question naturally arises whether
there are isochronous potentials defined by entire functions?
Another still  open and interesting question is
the investigation of the isochronicity property of non-monotonic
potentials.

The second author thanks
 the \'Sniadecki Foundation (Poland) and the Foundation of Fundamental
 Research of the Republic of Belarus for their support of the research
 and Professor M.Robnik for kind invitation to visit CAMTP and hospitality.

\end{document}